\documentclass[a4paper,11pt]{article}
\pdfoutput=1 

\usepackage{jinstpub} 
\usepackage{graphicx}
\DeclareGraphicsExtensions{.jpg,.pdf,.png}

\title{Fast component re-emission in Xe-doped liquid argon}

\author[a,b]{D. Akimov,}
\author[a,b]{V. Belov,}
\author[a,b,c]{A. Konovalov,}
\author[b]{A. Kumpan,}
\author[b]{O. Razuvaeva,}
\author[a,b,1]{D. Rudik,\note{Corresponding author.}}
\author[a,b,c]{and G. Simakov}

\affiliation[a]{Institute for Theoretical and Experimental Physics named by A. I. Alikhanov of National Research Centre "Kurchatov Institute"\\Moscow, 117218, Russian Federation.}
\affiliation[b]{National Research Nuclear University MEPhI (Moscow Engineering Physics Institute)\\ Moscow 115409, Russian Federation.}
\affiliation[c]{Moscow Institute of Physics and Technology (State University)\\Dolgoprudnyi, Moscow Region, 141700, Russian Federation}

\emailAdd{rudik.dmitry@mail.ru}

\abstract{We present the first direct experimental confirmation of the fast component re-emission in liquid argon (LAr) doped with xenon (Xe). This effect was studied at various Xe concentrations up to $\sim$3000 ppm. The rate constant of energy transfer for the fast component was quantified. It was shown that LAr doped with a high concentration of Xe without TPB has a better Pulse Shape Discrimination (PSD) efficiency than pure LAr or Xe-doped LAr with TPB. The stability of LAr+Xe mixture was tested for the first time at high Xe concentration for long continuous runtimes.}


\begin{document}
\maketitle
\flushbottom

\section{Introduction}
\label{sec:Intro}

Liquid Argon (LAr) is used as a detection medium in various physics experiments such as Dark Matter searches~\cite{martoff2017darkside, rubbia2006ardm}, neutrino experiments~\cite{brailsford2018dune, tayloe2018cenns} and other applications~\cite{lopes2003liquid}. It is cost-effective compared to Xe, for example, allowing the construction of kiloton detectors, can be easily purified, and has a scintillation efficiency comparable with other liquid noble gases~\cite{nobleGasDetectors}.  As a scintillator, the LAr provides very efficient Pulse Shape Discrimination (PSD) between different types of particles due to the significant difference in relative intensities of the fast and slow components~\cite{kubota1982liquid}. However, a significant disadvantage of LAr is that the wavelength of scintillation  lies in the VUV range ($\sim$128 nm). A common and convenient solution of this problem is to use wavelength shifters (WLS).

For instance, the COHERENT collaboration~\cite{akimov2017observation, akimov2018coherent} is currently taking data with the LAr CENNS-10 detector and is planning to deploy a ton-scale LAr detector in the near future. The signal readout is performed by a combination of visual light sensitive PMTs and WLS \cite{tayloe2018cenns}. Thus, the WLS optimisation for re-emission of the LAr scintillation light is very important in order to improve the detector response.

The most commonly used WLS for LAr is tetraphenyl butadiene (TPB). It is used for coating PMTs, detector walls and other elements of optical systems \cite{burton1973fluorescence, francini2013vuv}. The main disadvantage of any WLS film is low geometrical efficiency, since the light is re-emitted in a $4\pi$ solid angle, sensitivity to mechanical stress, long term stability~\cite{asaadi2019emanation}, scattering and re-absorption of the re-emitted light inside the WLS layer~\cite{francini2013vuv, stolp2016estimation} and also dependence of the WLS efficiency on the coating method.

An elegant idea is to use volume-distributed WLS, which can provide a higher efficiency of re-emission and better collection of the scintillation light. This improves position reconstruction since the re-emission occurs in the point of interaction.

It was shown that gaseous xenon doped in LAr works as a volume-distributed WLS shifting the wave length from 128 nm to around 175 nm~\cite{cheshnovsky1972emission, kubota1993suppression, minerskjold1993investigation}. Currently several groups are studying its wavelength shifting parameters and the properties of the LAr+Xe mixture~ \cite{neumeier2015intense, buzulutskov2017photon, peiffer2008pulse, wahl2014pulse}. As shown previously~\cite{akimov2017study}, Xe doped in small concentrations (up to 260 ppm by mass) re-emits only the slow component of LAr scintillation. Thus, at small concentrations it is impossible to use Xe as a single stage WLS and to keep PSD capability of the LAr+Xe mixture at the same time.

In previous studies with a broad concentration range (up to 1000 ppm by mass) \cite{peiffer2008pulse, wahl2014pulse}, shown that Xe-doping slightly improves the light yield and the energy resolution. Also they demonstrated that the PSD capability degrades with decreasing of Xe concentration in LAr. Both experiments \cite{peiffer2008pulse, wahl2014pulse} have shown that at the concentrations of Xe > 300 ppm the PSD capability of the LAr+Xe mixture is higher than that of pure LAr.

One experimental study~\cite{wahl2014pulse} indicated evidence of fast component re-emission at high Xe concentrations. However, the measurement scheme was very complicated, statistics were quite low, and the effect of the re-emission by Xe was obscured by the use of TPB. In our study, we confirm the clear observation of the fast component re-emission in LAr doped at high Xe concentration ($\sim$1100 ppm by mass). We show the dependence of this effect on the increase of Xe concentration and its relation to PSD efficiency. We also present the first experimental measurement of the energy transfer constant for the fast component of LAr scintillation.

\section{Scintillation}
\label{sec:LightEmission}
\subsection{Pure LAr}

The processes that follow ionizing particle interactions with liquid noble gases are well described e.g. in~\cite{nobleGasDetectors}. Creation of an excited Ar molecule and its radiative decay may be expressed by the following reactions (formulas 2.3 and 2.4 from~\cite{nobleGasDetectors}):
\begin{equation}
    \label{eq:Ar_decay}
    \begin{split}
        &\mathrm{Ar^* + 2Ar} \rightarrow \mathrm{Ar^*_2 + Ar}\\
        &\mathrm{Ar^*_2} \rightarrow \mathrm{2Ar + h\nu(128\ nm)}
    \end{split}
\end{equation}
During ionization-recombination and excitation processes, singlet $\mathrm{(^1\Sigma_u^+)}$ and triplet $\mathrm{(^3\Sigma_u^+)}$ excited states of $\mathrm{Ar_2^*}$ molecules are formed. The proportion between them depends on the ionization density, and thus, on the type of ionizing particle. Lifetimes of these states are significantly different: $\sim$7 and $\sim$1600 ns for the singlet and triplet states, respectively~\cite{kubota1982liquid}. Thus, the intensity of scintillation in pure LAr can be described as follows:
\begin{equation}
\label{eq:light_emission_in_pure_LAr}
    r = I_{1} e^{-t/T_f} + I_{2} e^{-t/T_s},
\end{equation}
where $T_f$ and $T_s$ are the decay times for singlet (fast component) and triplet (slow component) excited states, respectively. $I_1$ and $I_2$ are the intensities of these terms. A PSD analysis is possible by calculating the ratio of the light fraction in the first several tens of ns to the total detected light \cite{lippincott2008scintillation}. 
The capability to reduce electron recoil background using PSD analysis is one of the significant advantages of LAr for the experiments, where detection of nuclear recoils is required. For most applications, it is very important to keep this capability for xenon-doped LAr.

\subsection{LAr doped with Xe}

It is well-known that Xe-dopant works as a WLS in LAr \cite{cheshnovsky1972emission}. The widely used mechanism of energy transfer from $\mathrm{Ar_2^*}$ to Xe was proposed by S.~Kubota et al in 1993~\cite{kubota1993suppression}:
\begin{equation}
\label{eq:Kubota_model}
	\begin{split}
	    &\mathrm{Ar_2^*(^{1,3}\Sigma_u^+)} \rightarrow \mathrm{Ar_2^*(^{1,3}\Sigma_g^+) + h\nu(128\ nm)}\\
		&\mathrm{Ar_2^*(^3\Sigma_u^+) + Xe + (migration)} \rightarrow \mathrm{(ArXe)^* + Ar} \\
		&\mathrm{(ArXe)^* + Xe + (migration)} \rightarrow \mathrm{Xe_2^*(^{1,3}\Sigma_u^+) + Ar} \\
		&\mathrm{Xe_2^*(^{1,3}\Sigma_u^+)} \rightarrow \mathrm{2Xe + h\nu (175\ nm)}
	\end{split}
\end{equation}
This mechanism was proposed to transfer excitation only for the slow decay component of LAr scintillation, while the singlet state was expected to decay too fast for a transfer to happen, so it can only emit the 128 nm photon according to the first line in formula~\eqref{eq:Kubota_model}. Only in 2014, C.G. Wahl et al. introduced the possibility for singlet states to transfer their energy with the same mechanism and with the same rate constant~\cite{wahl2014pulse}. Thus, the intensity of the LAr+Xe mixture scintillation may be described as follows:
\begin{equation}
\label{eq:Wahl_model}
	r = I_{1} e^{-t/T_f} + I_{2} e^{-t/T_s} - I_{3} e^{-t/T_d},
\end{equation}
where $T_f$ and $T_s$ are the decay times for fast and slow component respectively, $T_d$ is the energy transfer time, equal for the singlet and triplet states. $I_1$, $I_2$ and $I_3$ are the intensities of these three terms, respectively.

However, it is not obvious that the rate constants for the singlet and triplet states should be equal. For example, the author of the overview \cite{buzulutskov2017photon} of the Xe-doped LAr experiments cited different values for the rate constants for the singlet and triplet states based on A.  Hitachi's \cite{hitachi1993photon} theoretical prediction. According to A. Hitachi the concentration of Xe in LAr should be much higher (about 200 ppm) to start the energy transfer process for the fast component. At the same time, according to \cite{hitachi1993photon} the value of the rate constant, which is proportional to the inverse transfer time, is higher for the singlet state than for the triplet one. Thus, the energy transfer time for the singlet state should be lower then that for the triplet state.

Taking this into account, we propose an extension of the model \eqref{eq:Wahl_model} for the high Xe concentrations with the fourth term which describes the energy transfer for the fast component:
 \begin{equation}
\label{eq:our_model}
	r = I_{1} e^{-t/T_f} + I_{2} e^{-t/T_s} - I_{3} e^{-t/T_{df} }- I_{4} e^{-t/T_{ds} },
\end{equation}
where $T_{df}$ and $T_{ds}$ represent energy transfer times for the fast and slow components separately.

\section{Experiment}

\subsection{Test chamber}
\label{sec:TestChamber}
\begin{figure}[tbp]
\centering
\includegraphics[width=.8\textwidth]{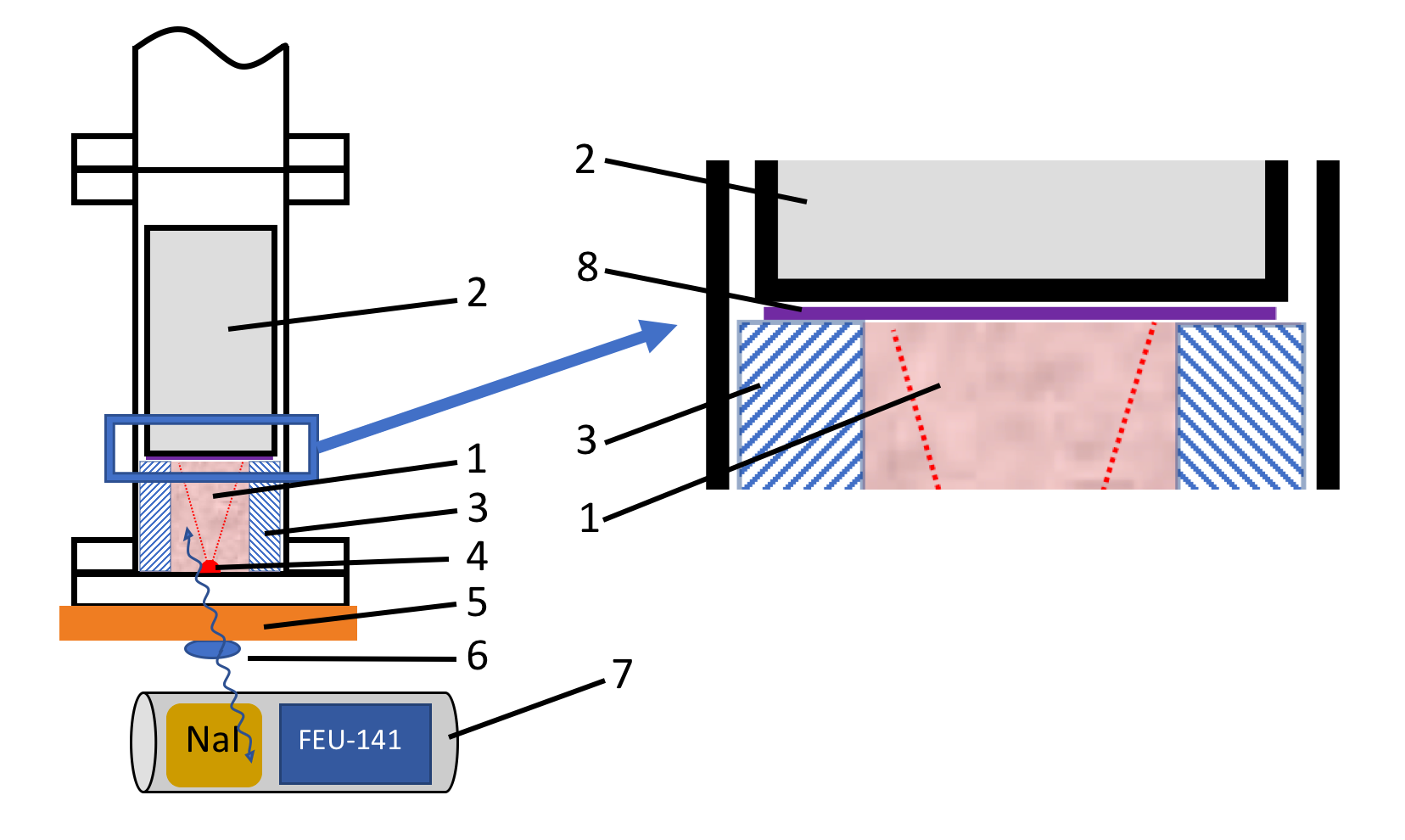}
\caption{\label{fig:test_chamber} Schematic view of the cold part of the test chamber (cryostat is not shown). 1 --- inner volume; 2 --- PMT FEU-181; 3 --- Teflon insert; 4 --- $\alpha$-source $^{241}$Am; 5 --- bottom part of the copper housing; 6 --- $\gamma$-source $^{22}$Na; 7 --- NaI[Tl] detector; 8 --- FS filter.}
\end{figure}

A schematic view of the cold part of the test chamber used for the experimental study of LAr doped with Xe is shown in the figure \ref{fig:test_chamber}, see also refs. \cite{akimov2017study, akimov2017new}. The test chamber was constructed on the base of a standard stainless steel vacuum 2.75" Conflat Flange (35 mm inner diameter). A Teflon insert, 22 mm inner diameter and 33 mm height (3), was intended to increase the light collection by reflection of the re-emitted light from its walls. The inner volume (1) is viewed by a multialkali FEU-181 PMT (2) produced by MELZ (Moscow). The FEU-181 has a MgF$_2$ window transparent to the VUV light down to 115 nm and therefore it is sensitive to the pure LAr scintillation light. A fused silica (FS) filter (8) with a wavelength cutoff at around 160 nm was used to eliminate the direct LAr light and to observe only the re-emitted light. Instead of a FS filter one can also place different optically transparent samples coated with a WLS film or detect the LAr scintillation light directly without the samples and the filter. 

Both $\alpha$- and $\gamma$-sources ($^{241}$Am (4) and $^{22}$Na (6)) were used in our tests. The $\alpha$-source was attached with glue to the middle of the bottom flange inside the working volume. The $\gamma$-source was placed beneath the bottom flange and copper housing (5). Since the $^{22}$Na source emits two 511-keV $\gamma$-rays in opposite directions \cite{page1957measurement} it was possible to use a coincidence scheme with a NaI[Tl] detector (7) placed beneath the test chamber for event tagging. 

A quick pulse height analysis was performed with the use of ORTEC Model 570 Spectroscopy Amplifier and ORTEC Model 927 Multichannel Analyser (MCA) (with original software). For event-by-event analysis, a digital 4-channel oscilloscope Tektronix TDS 5034 was used. One channel was used for recording the signals from the test chamber PMT and the other one for the signals from the NaI[Tl] detector in coincidence with the test chamber PMT. In each channel, the event sampling rate was 625 MS/s, and the full event length was 2500 samples (1.6 ns per sample). An amplitude threshold trigger on the test chamber PMT signal was used for signal recording.

\subsection{Mixture preparation}
\label{sec:MixturePreparation}
The Ar+Xe mixture was prepared at room temperature. This was done in order to avoid possible problems with Xe freezing on the detector walls in the case when Xe is added separately to LAr. The gaseous Ar was stored in a high pressure cylinder with a volume of $1230 \pm 50\: \mathrm{cm^3}$. The amount of Ar in the cylinder was estimated from pressure measurements. The evacuated small-volume pipeline ($180 \pm 5\: \mathrm{cm^3}$) of the gas system was filled with a small amount of xenon. The exact value was controlled by a precise mechanical vacuum-meter. Then the valve between the cylinder and the pipeline was opened. The calculated concentration of Xe in Ar in terms of mol/mol is based on the known amounts of gases in these volumes. The mixture with higher Xe concentration was prepared by adding a new portion of Xe to the existing mixture according to the described procedure. In this paper, the xenon concentration is presented in terms of gram/gram as used in previous studies by other groups \cite{neumeier2015intense, wahl2014pulse}. The purification of the Ar or Ar+Xe mixture was performed in the gas phase during the liquefaction process using a Mykrolis\footnote{\url{www.mykrolis.com}} gas filter.

The accuracy of the mixture prepared in this procedure is not very high, especially after several sequential additions of Xe. Given the uncertainties of volume and pressure measurements, the relative error of Xe concentration in each individual addition was 10 to 30\% depending on the amount of Xe in the addition. After several consecutive additions the total relative error of Xe concentration could reach 50\%. In order to increase the precision of this process, the independent measurements of Xe concentration in two different Ar+Xe mixtures were done with a GC-HRT chromatography mass-spectrometry system LECO Pegasus, which consists of gaseous chromatograph Agilent Technologies 7890B and High Resolution TOFMS. These measurements were performed by the Laboratory of Physical and Chemical Research of Federal State Unitary Enterprise Research and Technical Center of Radiation-Chemical Safety and Hygiene (FSE RTC RCSH). Taking these measurements as reference points, we reduced the error on Xenon concentrations to $\sim$10\%.

It was shown in ref. \cite{yunker1960solubility} that Xe is soluble in LAr at $\sim$87K up to 16\% by weight without any problems. However, a possibility of Xe aggregation in clusters of ice at this temperature was reported in ref. \cite{raz1970experimental}. To reduce the likelihood of this problem, all tests were done at higher temperature, $\sim$94 K \cite{lemmon2011nist}. In order to detect possible icing, a long continuous run was performed, and the stability monitored. The result of this test is presented in section \ref{sec:MixtureStability}.

Due to the significantly different partial pressure of Xe and Ar in the mixture one may expect different Xe concentration in the mixture after liquefaction with respect to the initial Xe concentration. This effect may be significant only if part of the prepared mixture is condensed \cite{neumeier2015optical}. It is negligible if the entire mass of mixture is condensed. At the same time, in our setup the mixture parameters can be controlled indirectly during the liquefaction process by checking the alpha peak position with the online monitoring system. Since the light yield depends on the Xe concentration, the alpha peak should change its position with increasing or decreasing Xe concentration.

\subsection{Measurement procedure}
Each dataset for each test chamber configuration and Xe concentration was collected during separate runs. The full measurement procedure can be described as the following steps.
\begin{figure}[tbp]
    \centering
    \includegraphics[width=\textwidth]{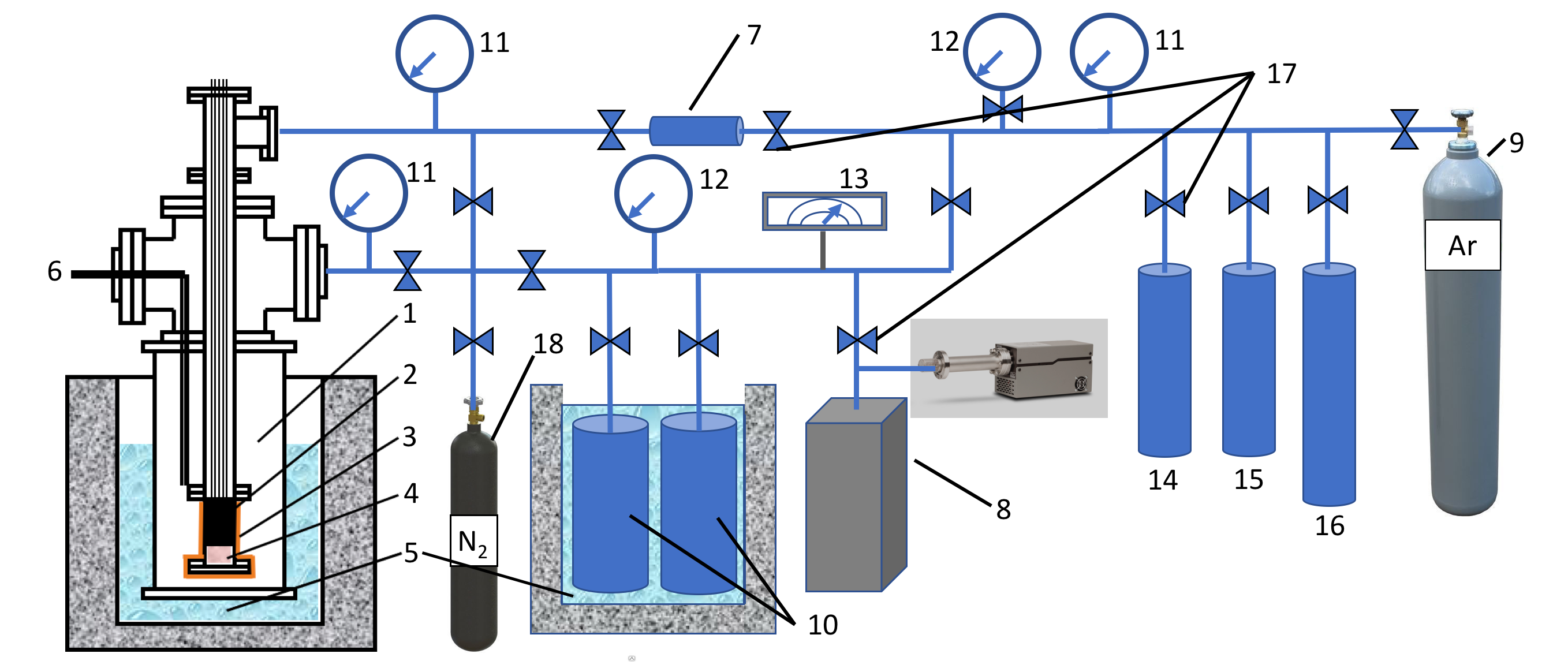}
    \caption{Test chamber gas system: 1 --- vacuum vessel; 2 --- PMT; 3 --- copper housing with a wire heater attached; 4 --- inner volume; 5 --- LN$_2$ in thermally insulated volumes; 6 --- thermocouple and electric heater wires; 7 --- Mykrolis gas filter; 8 --- magnetic discharge pump NORD-250 assembled with RGA; 9 --- gas cylinder with high pure Ar (class 5); 10 --- cryosorption pumps immersed into the LN$_2$; 11 --- mechanical manometers; 12 --- precise mechanical vacuum-meters; 13 --- ionization-thermocouple vacuum gauge VIT-2; 14--16 --- small cylinders with pure Xe, pure Ar and Ar+Xe mixture respectively; 17 --- valves; 18 --- gas cylinder with N$_2$.}
    \label{fig:gas_system}
\end{figure}

\begin{enumerate}
    \item\label{it:reassembling} Assemble the test chamber if it was opened to change internal components, e.g. for replacing the FS filter with the TPB sample.
    \item\label{it:pumping} Evacuate the gas system (figure~\ref{fig:gas_system}) and inner volume of the test chamber with a 3-stage pumping procedure with cryosorption (10) and magnetic discharge (8) pumps. First stage is the initial pumping of the gas system down to $10^{-1}$ Torr with a coal filled cylinder cooled with liquid nitrogen (LN$_2$). Second stage allows a reduction in pressure down to around $10^{-4}$ Torr with zeolite filled cylinder also cooled with LN$_2$. The final stage is a fine pumping with a magnetic discharge pump NORD-250 down to around $10^{-6}$ Torr. Vacuum control was carried out by means of the VIT-2 ionization-thermocouple vacuum gauge (13).
    \item\label{it:leak_test} Perform a leak test with the RGA (8) using its original software. The RGA is also used for vacuum control and gas analysis at the last stage of pumping process.
    \item Check the PMT dark rate test in order to confirm that PMT is working and stable.
    \item Prepare the Ar+Xe mixture as described in section~\ref{sec:MixturePreparation}.
    \item\label{it:liquifaction} Liquefy Ar or Ar+Xe mixture into the inner volume (4) of test chamber. LN$_2$ is filled into the surrounding thermally insulated bath (5). The cooling power was controlled by the pressure in the N$_2$-filled vacuum vessel (1).
    \item Reach thermodynamic equilibrium and desired temperature by means of a combination of cooling power of LN$_2$ and tuning power of the heater  wound around the copper housing (3).
    \item Record data using the oscilloscope as described in section~\ref{sec:TestChamber}. 
    \item\label{it:evacuation} Evacuate the gas from the inner volume at the end of the run. An appropriate gas storage cylinder (14--16) is immersed into LN$_2$ to cool it down. LN$_2$ is removed from the thermally insulated bath surround the vacuum vessel (5). Wire heater is kept on until the temperature of inner volume becomes larger than the Xe boiling point (around 166 K), to ensure that all Xe was evaporated and condensed into the storage volume. Argon vapor pressure at the boiling temperature of LN$_2$ (around 77.4 K) is around 0.26 bar~\cite{liquidegas} while xenon vapor pressure is negligible. Thus, during the evacuation process a small amount of Ar remained in the test chamber and the mixture compound was slightly changed after each test.
    \item Warm up the test chamber to room temperature.
\end{enumerate}

In order to check the stability of the entire system after reassembling the test chamber, we performed occasional engineering runs with pure LAr. We measured the base parameters of argon scintillation in these tests, such as the $\alpha$-peak position and slow component decay time, and used it to check the detector performance and to correct for parameter shift introduced by the reassembling process. Such comparison was not possible with the FS filter installed, since FS is almost completely opaque to the scintillation light in this case (see section~\ref{sec:MainResult}). At the same time, due to details of the evacuation procedure~(\ref{it:evacuation}), it is unwise to use Ar+Xe mixture for this purpose since this would change Xe concentration after each run. Therefore, we assume that detector reassemble effects are the same for the runs with the FS filter as for every other tests. Thus, overall systematic uncertainty between different runs was estimated to be at the level of 4\% as a maximum shift of the base parameters in the same configuration of the test chamber.

As mentioned above, the Xe concentration in the Ar+Xe mixture may change during the gas evacuation procedure. This effect was taken into account when Xe concentrations were calculated. Using the independent Xe-concentration spectrometric measurements, the influence of this effect was eliminated.

\subsection{Analysis}

We use two independent approaches to analyze the recorded waveforms. Both approaches are based on pulse finding algorithms. The first one uses the integration window method described in ref. \cite{akashi2015improving}: the integration window of 5 bins width is moving through the waveform bin by bin and integrate amplitudes within its bounds; when the integral crosses threshold level, the middle of integration window defines the start of the pulse; when integral crossing the threshold level in opposite direction, the middle of integration window is denoted as a last bin of the pulse. The second approach uses a two-threshold algorithm, which works in the following way. For each waveform, the amplitudes of all samples were histogrammed (projected on a vertical axis). This histogram was fitted by a Gaussian with $\sigma$ as rms since the majority of the waveform samples belong to a baseline which is spread out because of random noise. The mean value of the Gaussian was used as a baseline position. The first threshold was set at a level of $n\sigma$ to find pulses associated with signals produced by scintillation photons in the PMT; the second threshold $k\sigma$ ($k<n$) was set just above the baseline noise in order to determine the boundaries for integration of the pulse with minimal area losses. The parameters $n$ and $k$ (equal to 4 and 2 in our case respectively) were obtained from the analysis of single photoelectrons (SPE) on the signal tails. In both approaches, the SPE pulse height distributions were obtained from the scintillation signal tails.

Both methods have shown very similar results for the light yield (LY) and the PSD capability versus Xe concentration. Further analysis was performed for the events that passed the selection criteria described in \ref{sec:ss_selection}. Our study was focused primarily on the behavior of the averaged waveforms for $\alpha$-particles and the capability of PSD between $\alpha$ and $\gamma$ bands with the increase of Xe concentration. These two characteristics are most sensitive to the Xe concentration and demonstrate very clearly the VUV light re-emission by Xe-doping (which was shown in previous studies and is presented in section~\ref{sec:MainResult}).
The non-uniformity of the test chamber light response did not let us observe a 511-keV $\gamma$-peak from the $\mathrm{^{22}Na}$ source. The $\gamma$-events were used for the PSD capability investigation only.

In our PSD analysis we used the so-called ``fraction $N$'' discrimination parameter (F$N$), which is defined as a fraction of the event area integrated within the first $N$ ns of the signal divided to the total event area. We used the parameter F40, differing from the common F90 parameter (with a 90 ns window) for the following reasons. In \cite{lippincott2008scintillation} it was pointed out, that although the 90 ns window provides the best separation capability, the variation of the window length from 50 to 250 ns did not substantially affect the discrimination capability. On the other hand, in another ref. \cite{peiffer2008pulse} shown that the best separation between different particle species was reached with the use of a time window between 35 and 45 ns. In our case, the shorter window is preferable because the triplet decay time decreases with the increase of Xe concentration as shown in \cite{kubota1993suppression, wahl2014pulse}. We also investigated the behavior of this discrimination parameter for different test chamber configurations and Xe concentrations as a function of width of the integration window. Results of this investigation are presented in section~\ref{sec:PSDDependence}

\subsection{Event selection}
\label{sec:ss_selection}

The following selection criteria (cuts) of events were used in this analysis.
First of all, events with pulses before the trigger were eliminated from further analysis in order to reduce the possibility of pileup (two or more events on the same waveform). However, this cut does not work for pileup at the end of waveforms. Secondly, events with very high amplitudes were eliminated in order to exclude PMT saturation. Thirdly, events with Cherenkov light and high-amplitude PMT noise were eliminated  by applying a cut on the number of pulses in the event. Both Cherenkov light and PMT noise have small number of pulses (less then 3 in our case). If averaged with other signals these events would distort the true shape of waveform, and thus, must be excluded.
\begin{figure}[tbp]
    \centering
    \includegraphics[width=.8\textwidth]{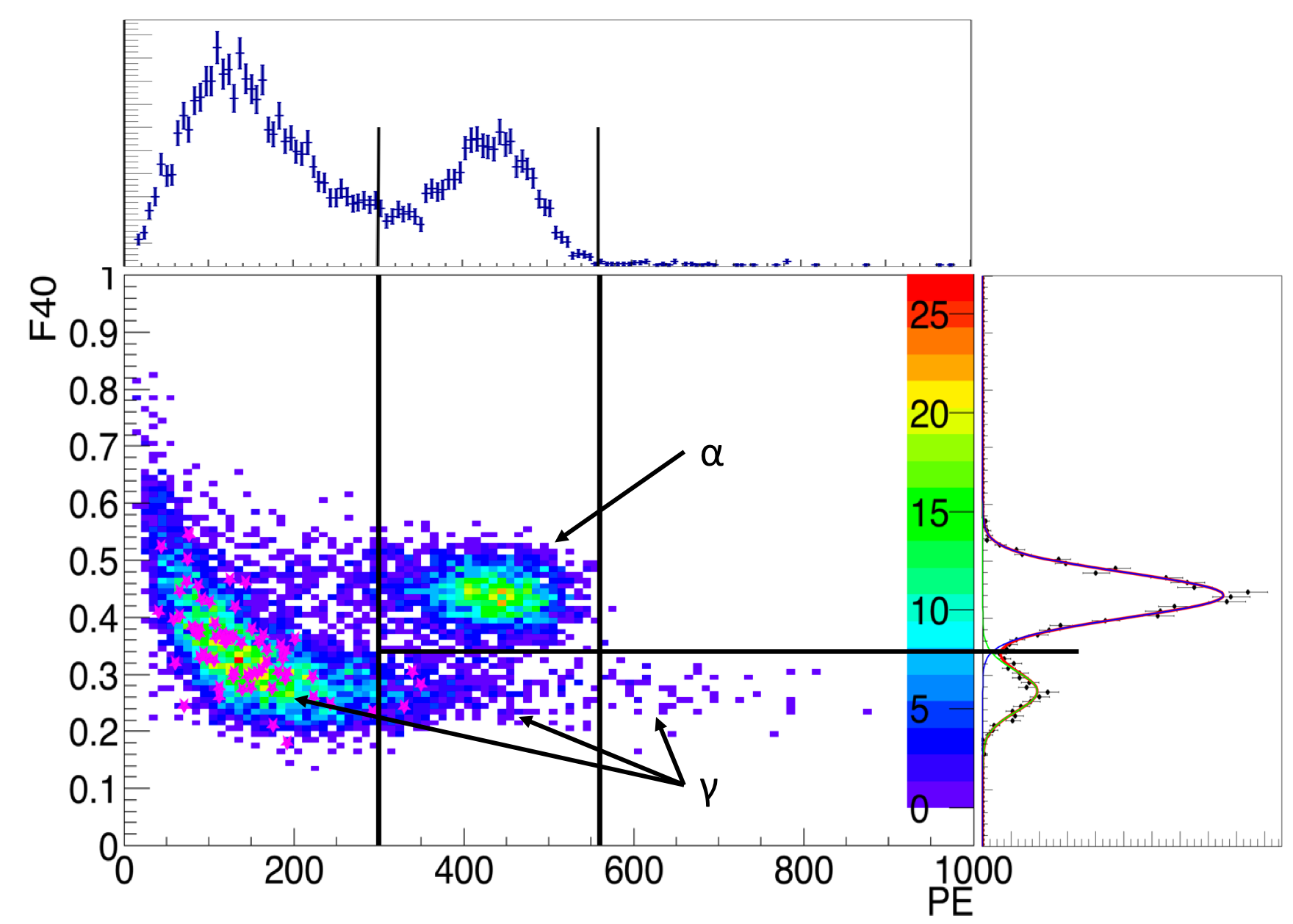}
    \caption{An example of a PSD plot with F40 vs number of PE (TPB sample combined with $\sim$2300 ppm Xe). The PE projection is shown above with the $\alpha$-source peak delineated by two lines. The F40 projection is shown to the right with the two Gaussian fits for $\alpha$ (blue line) and $\gamma$ (green line) bands; red line represents total fit. Magenta stars --- 511 keV $\gamma$-events selected with a NaI[Tl] signal in coincidence.}
    \label{fig:psd_parameter_example}
\end{figure}

To produce averaged waveforms the ``pure'' $\alpha$-events were selected from the alpha peak region, and as well, a PSD cut was used when possible. An example of the PSD plot with denoted alpha peak region and PSD parameter projection is presented in figure~\ref{fig:psd_parameter_example}. The PSD cut was chosen at the crossing point of Gaussian functions which were used for fitting PSD distribution in this case: for $\alpha$ (blue line) and $\gamma$ (green line) bands. ``Pure'' 511 keV $\gamma$-events selected with a NaI[Tl] signal in coincidence (magenta stars) was used to tag the $\gamma$-band. Note, 1274.5 keV $\gamma$-line of $^{22}$Na-source cannot be selected with the coincidence scheme but can produce high energy responses in the test chamber. For those datasets, in which the $\alpha$- and $\gamma$-events were indistinguishable by the PSD analysis,
a systematic uncertainty obtained from the characteristics of the $\gamma$-events within the same energy window as for alphas was added.

\section{Results}

\subsection{Re-emission of the fast component of LAr scintillation}
\label{sec:MainResult}

\begin{figure}[tbp]
\centering
\includegraphics[width=.8\textwidth]{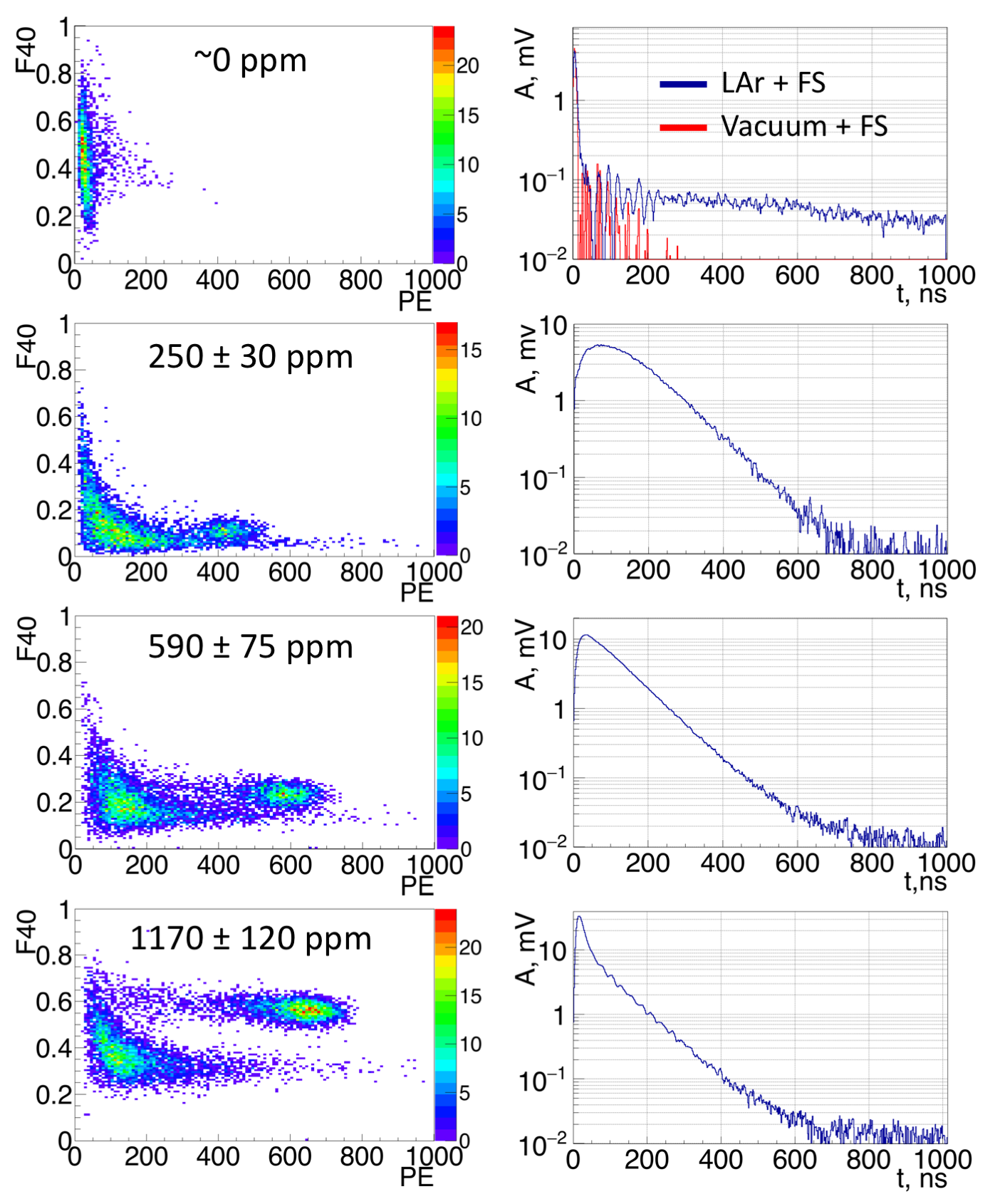}
\caption{\label{fig:main_result} 
Evolution of the PSD capability (F40 parameter versus signal area in PE; left column) and the averaged scintillation waveform shape (right column) of the LAr+Xe mixture with increasing Xe concentration for test with FS filter. Each row corresponds to the Xe concentration written on the scatter plot.}
\end{figure}

The main result of this study is the confirmation of the LAr scintillation fast component re-emission by the Xe-dopant (see the series of plots in figure~\ref{fig:main_result}).

The measurement with zero Xe concentration was performed in order to check the capability of the FS filter to cut the VUV scintillation light of LAr. Certain number of events were recorded in this test. Most were associated with PMT noise, as can be seen from a comparison of the averaged waveforms obtained with pure LAr and with the empty chamber (figure~\ref{fig:main_result}, right column, top line). In another fraction of events, there were pulses in a slow component region (see the same figure). These pulses may be caused by the re-emission of LAr scintillation by traces of Xe or $\mathrm{N_2}$ contamination ($\sim 1$ ppm by mass). The influence of these traces is negligible in the tests with high Xe concentration. 

For the low Xe concentrations (from 0 to $\sim$250 ppm by mass) there is no evidence of the fast component re-emission, as shown by us previously in ref. \cite{akimov2017study}. In the second row of figure~\ref{fig:main_result}, which corresponds to $250\pm30$ ppm Xe concentration, one can see that $\alpha$- and $\gamma$- events are almost indistinguishable in the PSD plot in the energy region, corresponding to the $\alpha$-source peak position. The shape of averaged waveform has a simple single-hump structure described by two processes: transfer of excitation from $\mathrm{Ar_2^*}$ to Xe atoms with formation of $\mathrm{Xe_2^*}$ excited molecules \eqref{eq:Kubota_model} and the subsequent decay of those. If the fast component of scintillation light is re-emitted by Xe at this concentration the shape of averaged waveform would be different from this one-hump structure: the fast component should manifest itself in the first several tens of ns which is not the case. Without re-emission of the fast component the model \eqref{eq:Wahl_model} is reduced to two terms:
\begin{equation}
    \label{eq:reduced_emission_model_without_fast_component}
    r = I_{1} e^{-t/T_s} - I_{2} e^{-t/T_{ds}},
\end{equation}
where $T_s$ and $T_{ds}$ are the decay and transfer times, respectively, for the slow component; $I_1$ and $I_2$ are the intensities of these terms.
Note, that there is an infrared emission from de-excitation of $\mathrm{(ArXe)^*}$ molecules described in papers \cite{neumeier2015intense, gerasimov2004optical}. However, this infrared component is invisible with our PMT. 

Further increase of Xe concentration to $590\pm75$ ppm (third line in the figure~\ref{fig:main_result}) resulted in sharper shape of the averaged waveform. Both $\alpha$- and $\gamma$-bands are shifted to higher F40 values, and separation between them in the energy region of the $\alpha$-source peak position became more clean (also, see figure \ref{fig:Qpsd_figure} in section \ref{sec:PSDDependence} for the quantitative efficiency of PSD). These two effects are the manifestation of the increasing re-emission of the fast component.

When the Xe concentration exceeds 1000 ppm (bottom row in the figure~\ref{fig:main_result}), one can see clearly two separate $\alpha$- and $\gamma$-bands on the scatter plot. The averaged waveform of $\alpha$-events no longer has single-hump structure, two different trends can be clearly distinguished. This two effects are direct manifestations of re-emission of both the fast and slow components. All 4 terms of the model \eqref{eq:our_model} should be used to describe light emission in this case: two terms for re-emission process for the fast and slow component and two terms for their subsequent decay. 

Further increase of Xe concentration did not change parameters significantly. Saturation was reached at $\sim$1700 ppm. We suppose that at this level of concentration all the LAr scintillation light is captured and re-emitted by Xe-dopant. That is in agreement with the spectrometric study performed in ref.~\cite{neumeier2015intense}.

\subsection{Scintillation waveform shape at high Xe-dopant concentration}
\label{sec:FastComponentHighXeConc}
\begin{figure}[tbp]
    \centering
    \includegraphics[width=.8\textwidth]{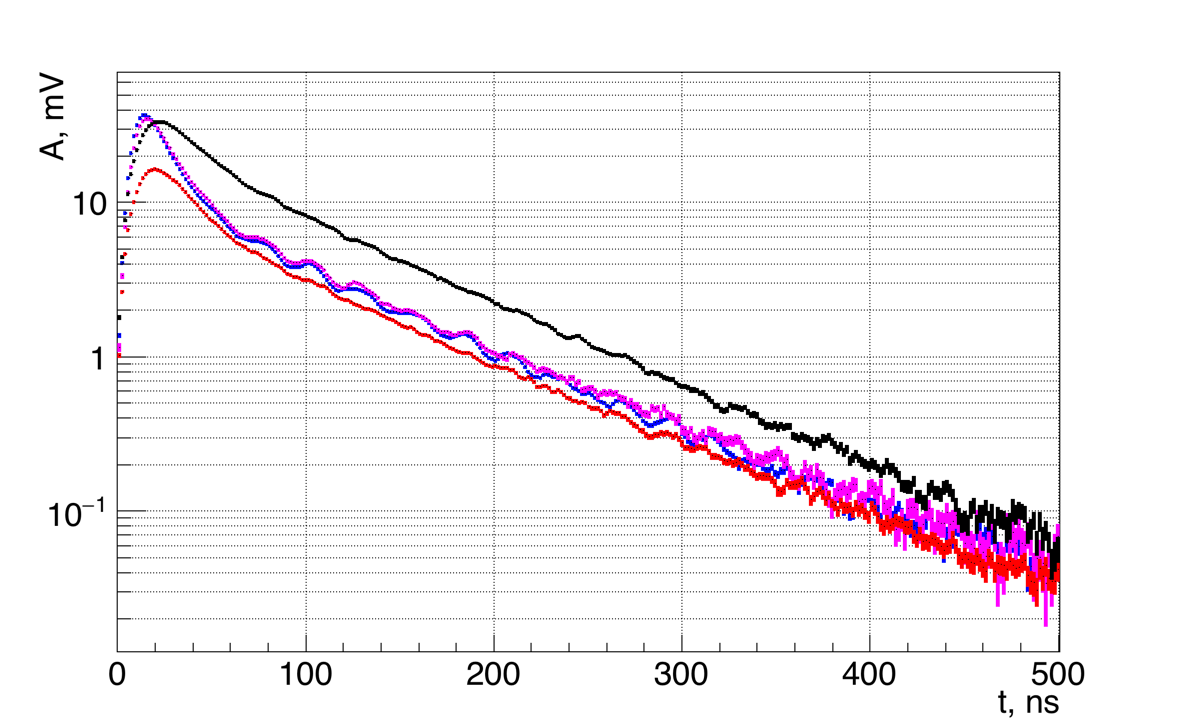}
    \caption{Averaged waveforms for different test chamber configurations and high Xe concentrations: no samples or filters, $\sim$2900 ppm Xe (black); FS filter with $\sim$1800 ppm Xe (blue) and $\sim$2900 ppm Xe (magenta); TPB sample, $\sim$2300 ppm Xe (red).}
    \label{fig:compare_averaged_wf}
\end{figure}
We did not carry out a detailed study with and without the FS filter and with TPB sample for every Xe concentration. However, a comparison study was done for these three configurations at the high Xe concentrations. The averaged waveforms obtained in this study are shown in figure~\ref{fig:compare_averaged_wf}. The blue and magenta lines are for runs with the FS filter at Xe concentrations of $\sim$1800 ppm and $\sim$2900 ppm, respectively. The shape of waveforms is practically the same. This reflects the fact that the reemission process has reached saturation at such concentrations. The other two waveforms, as red and black curves respectively, were obtained without the FS filter but with the TPB sample ($\sim$2300 ppm Xe) and with direct light detection ($\sim$2900 ppm Xe). One can see that both curves are very similar in shape: in both cases the light in the whole VUV region is detected, including Xe emission wavelength. The intensity of the detected light with TPB sample is lower than that in the direct measurement by a factor of $\sim$2.7. The main contributor to this decrease is an isotropic light re-emission in the TPB layer.

We expected to see the same shape of waveforms in the runs with and without the FS filter because of saturation of the pulse shape parameters at such Xe concentrations.  In such a case, all initial Ar excitations should be transferred to Xe and all photons emitted in the Xe emission band with wavelengths higher than the transparency cut-off of the FS filter. However, one can see that the shape of waveforms is quite different; the relative contribution of the slow component is larger when the FS filter is not used.

Such a difference in waveform shapes may be explained by the presence of the VUV component ($\sim$147 nm) of the excited $\mathrm{(ArXe)^*}$ molecule created during the migration process (see formula~\eqref{eq:Kubota_model}), although the study~\cite{neumeier2015intense} shows that for LAr+Xe mixture this component disappears at the high concentrations. Another study (see figure 14 in ref.~\cite{gerasimov2004optical}) shows the presence of this line in the gaseous mixture at Xe concentration in Ar up to 1\%. Further comprehensive studies are required to clarify this difference. 

\subsection{Scintillation signal parameters}
\label{sec:scintillation_parameters}

\begin{figure}[tbp]
\centering
\includegraphics[width=\textwidth]{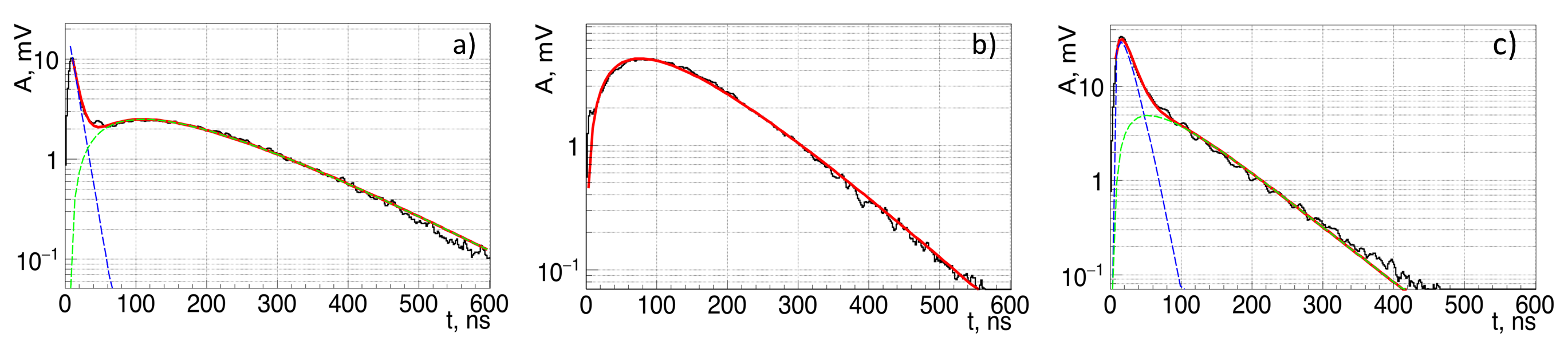}
\caption{\label{fig:fit_example} Fit examples for different Xe concentrations and measurement conditions: a) direct light measurement, $\sim$200 ppm Xe, 3-term fit model; b) FS filter, $\sim$250 ppm Xe, 2-term fit; c) FS filter, $\sim$1170 ppm Xe, 4-term fit model. Red solid line is total fit; blue and green dashed lines are fast and slow components, respectively; on (b) total fit is represented by the slow component only.}
\end{figure}

The scintillation signal parameters were measured for the full range of Xe concentrations including our previous results for the small Xe concentrations \cite{akimov2017study}. For this purpose, the averaged waveforms for the $\alpha$-events were approximated with the light emission models appropriate for different Xe concentrations (see Table \ref{tab:data_table} for details and figure \ref{fig:fit_example} for illustration). 
\begin{table}[htbp]
\centering
\caption{\label{tab:data_table} Conditions of tests and applied models; letters a, b, c denotes corresponding methods shown in figure \ref{fig:fit_example}}
\smallskip
\begin{tabular}{|c|c|c|c|}
\hline
Xe, ppm&Filter&Method&Formula\\
\hline
$0\pm1$ & -- & Two decay exponents & \eqref{eq:light_emission_in_pure_LAr}\\
$10\pm5$ & FS & (b) & \eqref{eq:reduced_emission_model_without_fast_component}\\
$70\pm20$ & TPB & (a) & \eqref{eq:Wahl_model}\\
$200\pm20$ & -- & (a) & \eqref{eq:Wahl_model}\\
$250\pm30$ & FS & (b) & \eqref{eq:reduced_emission_model_without_fast_component}\\
$400\pm40$ & FS & (b) & \eqref{eq:reduced_emission_model_without_fast_component}\\
$590\pm75$ & FS & (c) & \eqref{eq:our_model}\\
$1170\pm120$ & FS & (c) & \eqref{eq:our_model}\\
$1800\pm180$ & FS & (c) & \eqref{eq:our_model}\\
$2920\pm270$ & FS & (c) & \eqref{eq:our_model}\\
\hline
\end{tabular}
\end{table}

In the test with direct light measurements in pure LAr, simple fit by two decay exponents was used in accordance with the model~\eqref{eq:light_emission_in_pure_LAr}. At the low Xe concentrations, the fast component of LAr scintillation cannot be re-emitted by Xe, but is visible either with the TPB sample or in direct LAr+Xe scintillation light measurements (with PMT sensitive to 128 nm light). Therefore, a 3-term model \eqref{eq:Wahl_model} should be used to fit the averaged waveform in these two configurations (figure \ref{fig:fit_example}(a), red solid line). Contributions of the fast ($I_1e^{-t/T_f}$) and the slow ($I_2e^{-t/T_s}-I_3e^{-t/T_{ds}}$) components are shown by the blue and green dashed lines, respectively. 

In the tests with the FS filter and low Xe concentrations, the fast component cannot be visible, because the FS filter is opaque for the LAr light at $\sim$128 nm. A simplified two-term model \eqref{eq:reduced_emission_model_without_fast_component} should be used in this case to fit the averaged waveform (figure \ref{fig:fit_example}(b), red solid line). The fit is represented then by the terms associated with the slow component only ($I_1e^{-t/T_s}-I_2e^{-t/T_{ds}}$).  

At the high Xe concentrations (with the FS filter), a 4-term model \eqref{eq:our_model} should be used (figure \ref{fig:fit_example}(c), red solid line). Both the fast and slow components of LAr scintillation undergo transfer of excitation to Xe atoms with subsequent decay. Contributions of the fast ($I_1e^{-t/T_f}-I_3e^{-t/T_{df}}$) and the slow ($I_2e^{-t/T_s}-I_4e^{-t/T_{ds}}$) components are shown by the blue and green dashed lines, respectively.

Note, that the averaged signal rising in a first few ns depends not only on the excitation transfer mechanism but also on the PMT rise time and on the uncertainty in the selection of signal starting points when averaging. These effects were estimated and included into the systematic uncertainty of obtained results.

\begin{figure}[tbp]
\centering
\includegraphics[width=.45\textwidth]{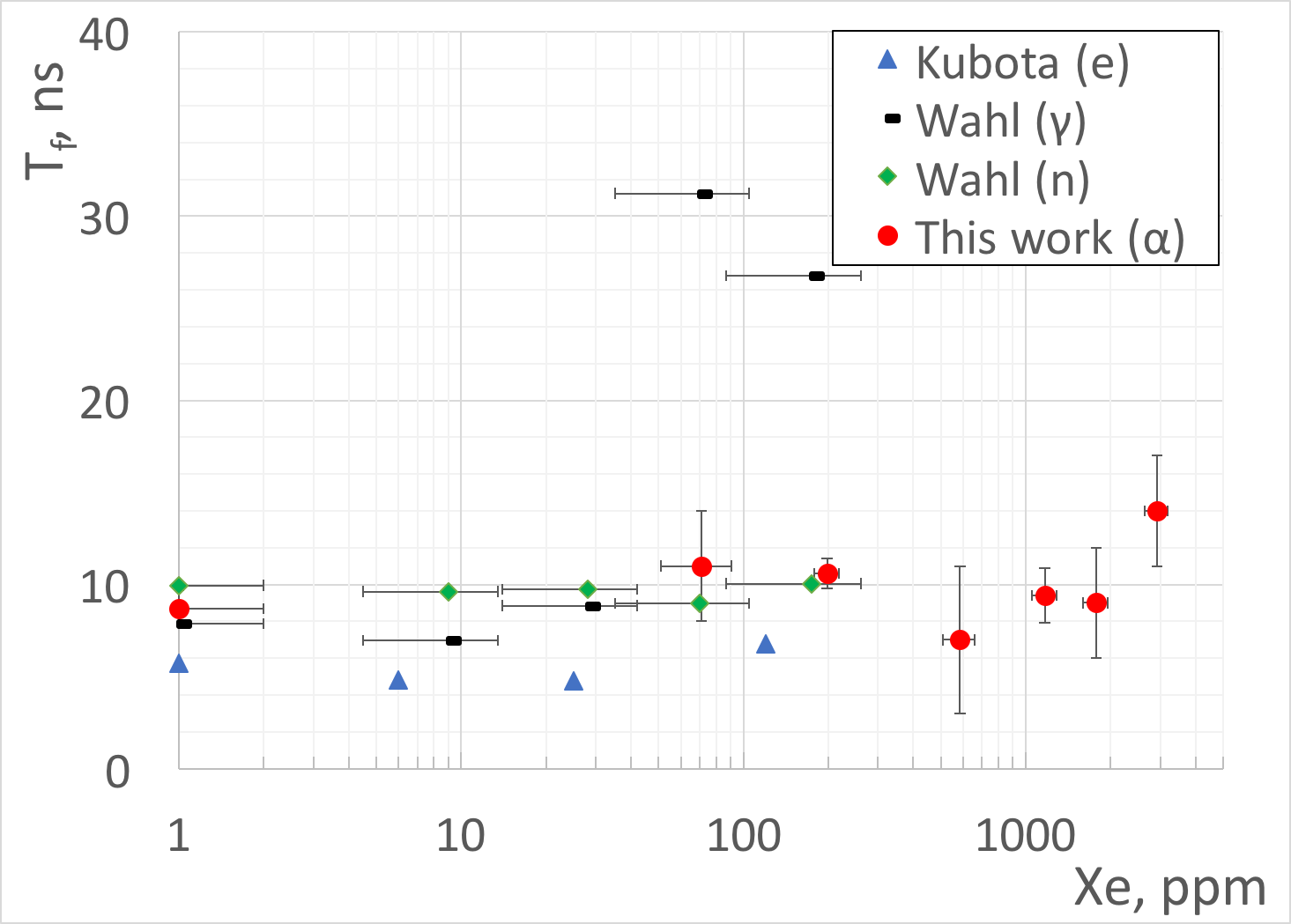}
\qquad
\includegraphics[width=.45\textwidth]{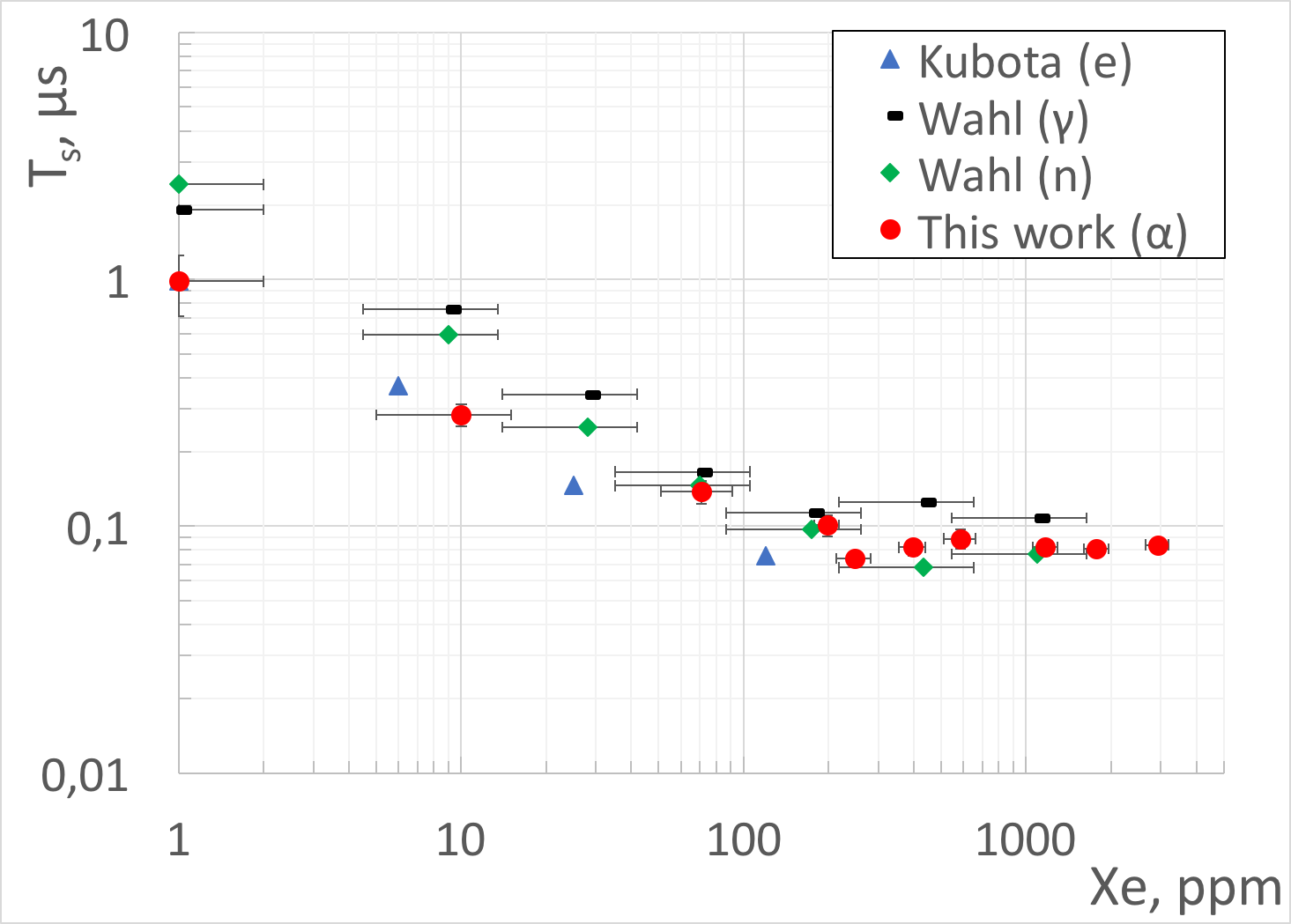}
\caption{\label{fig:fast_and_slow_comp} Fast (left) and slow (right) decay times for different Xe concentrations.}
\end{figure}

The decay time of the fast and slow components of scintillation versus Xe concentration is shown in figure \ref{fig:fast_and_slow_comp} together with previous results \cite{kubota1993suppression, wahl2014pulse}. For the fast component (figure \ref{fig:fast_and_slow_comp}, left), one can see that our measurements agree with other experiments which have data only in the range from 0 to 200 ppm Xe \cite{wahl2014pulse}. According to our data, the fast component decay time is almost the same for all points.

The most accurate measurements were done for the slow component decay time (figure \ref{fig:fast_and_slow_comp}, right). A good agreement with previous measurements was demonstrated in the full range of Xe concentration. A power law behavior in the range of concentration from 0 to $\sim$300 ppm is confirmed. We found also that for the higher levels of Xe concentration the slow component is constant at a level of about 85 ns.
\begin{figure}[tbp]
\centering
\includegraphics[width=.75\textwidth]{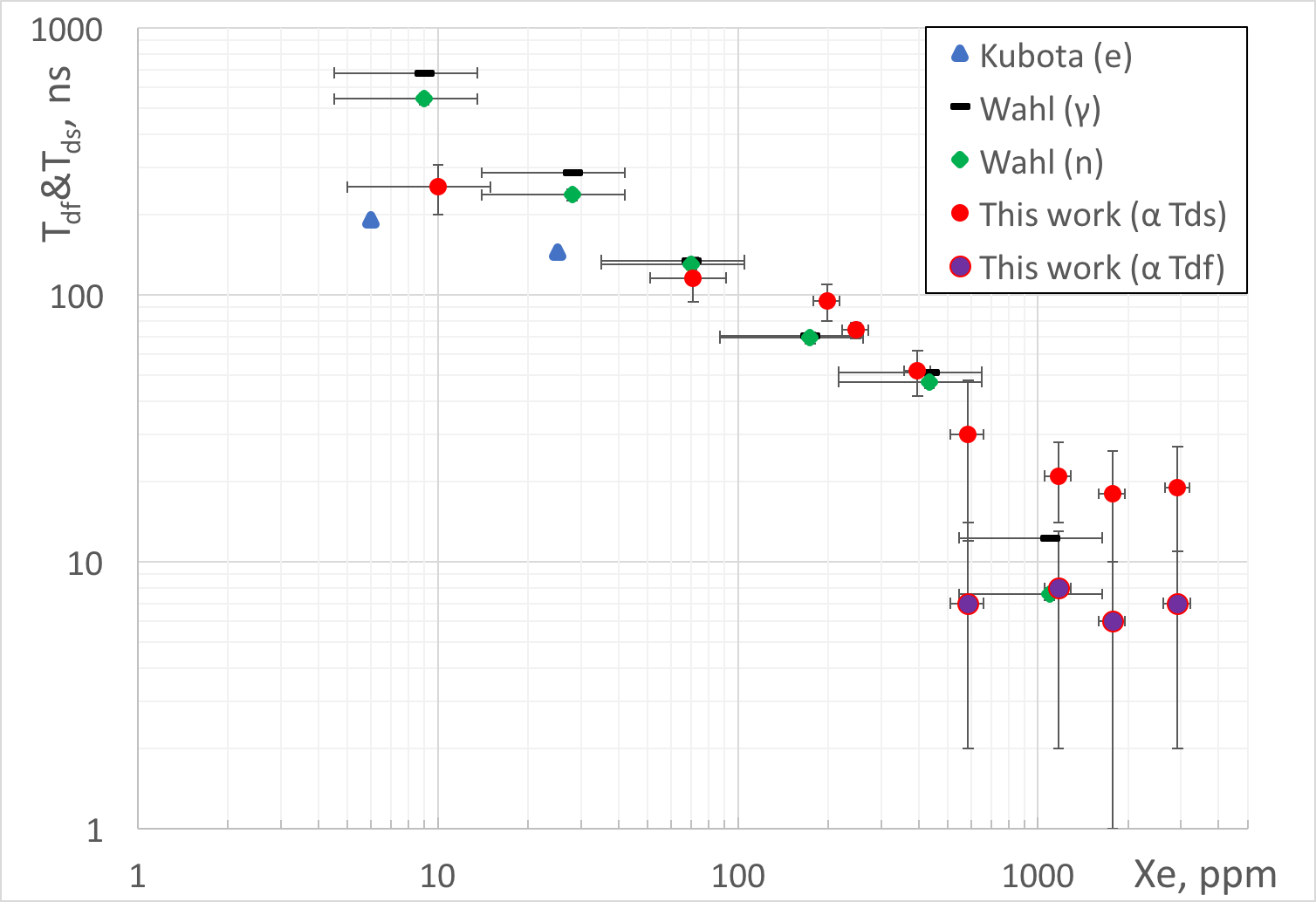}
\caption{\label{fig:transfer_time} Transfer time constants $\mathrm{T_{df}}$ and $\mathrm{T_{ds}}$ versus Xe concentration.}
\end{figure}

The dependence of transfer time constants $T_{df}$ and $T_{ds}$ versus Xe concentration is shown in figure \ref{fig:transfer_time}. Insufficient time resolution and presence of electronic noise resulted in quite large uncertainties in these parameters at high levels of Xe concentration. One can see from the plot that the introduction of the fourth term (with independent transfer time for the fast component $T_{df}$) into the light emission model allows $T_{ds}$ to follow a power law behavior up to the highest Xe concentrations. One may compare this result with that of ref. \cite{wahl2014pulse}, where the last points (for $\gamma$- and $n$-events) of $T_{ds}$ at $\approx$~1000 ppm of Xe are at least by about factor 2 lower then expected from a power law dependence.

For Xe concentration in the range 500 -- 1200 ppm, we have estimated from our data an energy transfer rate constant for the fast component  $k_{(\mathrm{^1\Sigma_u^+})}=\frac{1}{T_{d}\cdot[M]}=0.9^{+2.3}_{-0.31} \cdot 10^{-11} {cm^3}/{s}$, where $T_{d}$ is energy transfer time and $[M]$ is Xe concentration. This value is close within the errors to the one predicted by Hitachi \cite{hitachi1993photon}, $k_{(\mathrm{^1\Sigma_u^+})} = 3.3 \cdot 10^{-11} cm^3/s$. At the same time, the averaged $T_{df}$ value ($\sim$7 ns) appears to be equal to the fast  decay time component. In this case, the significant part of LAr singlet states would decay with emission of 128 nm photons before they transfer excitation to Xe. This part of the fast component would not be visible in the tests with the FS filter but should appear in the tests with TPB or with direct light detection, however this was not observed (see figure~\ref{fig:compare_averaged_wf}.  Also,  this contradicts the saturation effect reached at the high Xe concentrations.

Although the error bars of $T_{df}$ do not exclude the values of 1--2 ns (see figure~\ref{fig:transfer_time}) we checked two assumptions to explain the high $T_{df}$ value. First, systematic uncertainties of different origin, e.g. electronic noise or error in the signal start determination, can smear the beginning of the averaged waveform and increase the obtained $T_{df}$ value. We tried another signal start finding algorithm based on the threshold set at a quarter of maximal signal amplitude and did not find significant change of the waveform shape. Second, this might be a result of the incompleteness of the light emission model~\eqref{eq:our_model} based on~\eqref{eq:Wahl_model}, proposed by C.G Wahl et al in~\cite{wahl2014pulse}, where the terms related to the Xenon decay time were neglected due to the relatively short decay time for both fast and slow components of Xe scintillation (4 and 27 ns, respectively~\cite{nobleGasDetectors}). This approach might not be applicable for the high Xe concentration mixtures where the transfer time becomes comparable to the Xenon decay time. To evaluate this, an extended fit model with additional Xe decay terms was applied and no significant difference to 4-term fit was found. Therefore, the 4-term model can be used in our case. 

One speculative but possible explanation of this discrepancy is that the mechanism of energy transfer for the fast component can happen not only in a collision process but also through direct excitation of Xe by 128 nm photons:
\begin{equation}
\label{eq:new_energy_transfer_mechanism}
	\begin{split}
	    &\mathrm{Ar_2^*(^{1}\Sigma_u^+)} \rightarrow \mathrm{2Ar + h\nu(128\ nm)}\\
		&\mathrm{Xe + h\nu(128\ nm)} \rightarrow \mathrm{Xe^*} \\
		&\mathrm{Xe^* + Xe} \rightarrow \mathrm{Xe_2^*(^{1,3}\Sigma_u^+)} \\
		&\mathrm{Xe_2^*(^{1,3}\Sigma_u^+)} \rightarrow \mathrm{2Xe + h\nu (175\ nm)}
	\end{split}
\end{equation}
According to several studies~\cite{raz1970experimental,brunt1976study} this process is possible. Also, see the energy levels scheme of Ar and Xe excited states, for example in~ref.~\cite{buzulutskov2017photon}, figure 1. In this case, energy transfer from $\mathrm{Ar_2^*(^{1}\Sigma_u^+)}$ to $\mathrm{Xe_2^*(^{1,3}\Sigma_u^+)}$ has the same rate as the decay of $\mathrm{Ar_2^*(^{1}\Sigma_u^+)}$, therefore $T_{df}$ should be of the same value as the LAr fast component decay time and that is in agreement with our result.  This energy transfer mechanism for the fast component can also explain the relative enhancement of the slow component in the tests with TPB and in the direct measurements in comparison to those with the FS filter  (see figure~\ref{fig:compare_averaged_wf}). Indeed, if this mechanism is in effect, the mediated molecules $\mathrm{ArXe^*}$ can not be created for the fast component. Consequently, if there is a contribution from the decay of $\mathrm{ArXe^*}$ states only the slow component can increase without the FS filter.

\begin{figure}[tbp]
\centering
\includegraphics[width=.45\textwidth]{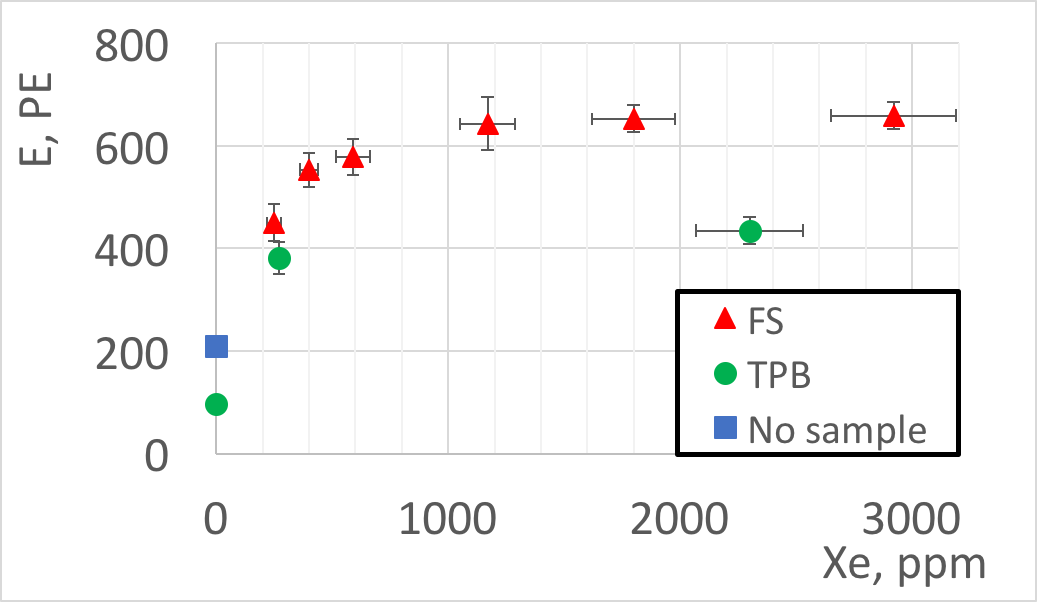}
\qquad
\includegraphics[width=.45\textwidth]{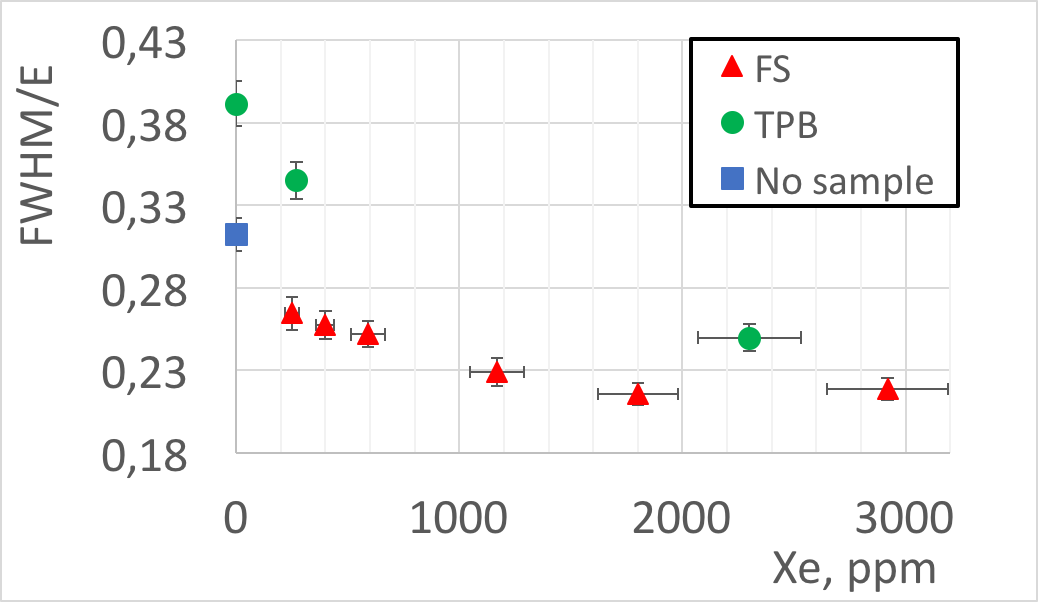}
\caption{\label{fig:ly_and_resolution_vs_conc} The dependence of $\alpha$-peak position (left) and its resolution (right) on Xe concentration; red triangles --- with FS filter; green circles --- with TPB sample; blue square --- test with no sample in front the PMT photocathode.}
\end{figure}
Other scintillation signal parameters are in agreement with the previous results \cite{peiffer2008pulse, wahl2014pulse, buzulutskov2017photon}: as Xe concentration grows, the light yield increases, and the energy resolution becomes slightly better (figure~\ref{fig:ly_and_resolution_vs_conc}). According to our data, the light yield saturates when Xe concentration reaches the point of the total re-emission of the fast component by Xe. 

\subsection{PSD dependence on Xe concentration}
\label{sec:PSDDependence}

In the tests with FS, we observed the PSD capability progressively improve (starting from almost no separation between $\alpha$- and $\gamma$-events) with increasing of Xe concentration. See examples of the PSD plots in figure~\ref{fig:main_result}.

To evaluate the quality of PSD we used two parameters. The first one ($Q_{PSD}$) was defined as the acceptance efficiency from $\alpha$-events for a PSD cut that provides 99.9\% rejections of $\gamma$-event based on the fit PSD distribution as shown in figure~\ref{fig:psd_parameter_example}. The second parameter was defined as $d=\frac{\mu_\alpha-\mu_\gamma}{\sqrt{\sigma_\alpha^2+\sigma_\gamma^2}}$, where $\mu_\alpha$ and $\mu_\gamma$ are the centers of ``fraction $N$'' discrimination parameter distribution for $\alpha$- and $\gamma$-bands located within an energy slice, corresponding to the $\alpha$-source peak position; $\sigma_\alpha$ and $\sigma_\gamma$ are their standard deviations~\cite{peiffer2008pulse} (see for example figure~\ref{fig:psd_parameter_example}). 

\begin{figure}[tbp]
    \centering
    \includegraphics[width=.8\textwidth]{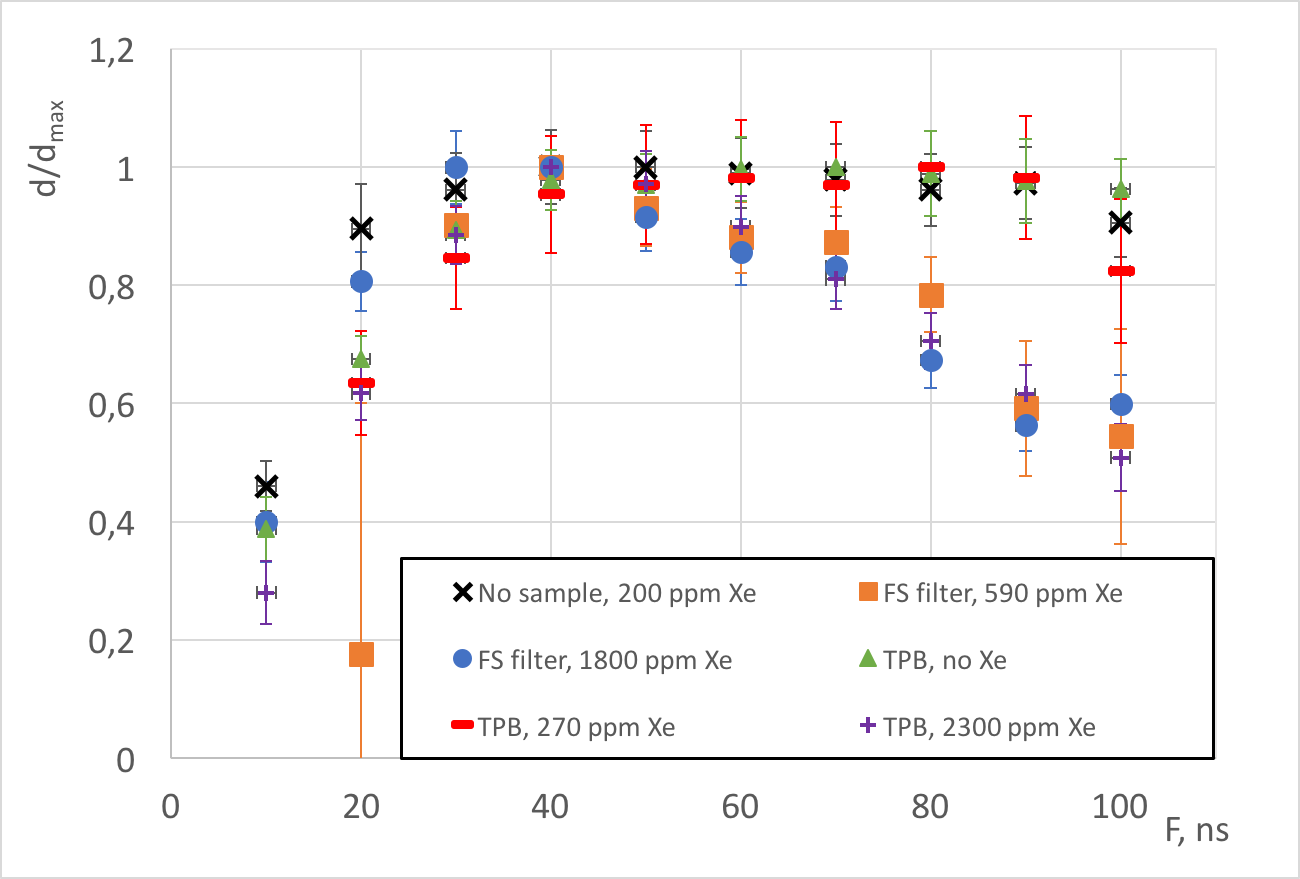}
    \caption{The dependence of $d/d_{max}$ ratio on fraction discrimination parameter F$N$ for the most important test chamber configurations and Xe concentrations, where $d_{max}$ is the maximal value of $d$ for all F$N$ in a given run.}
    \label{fig:psd_fractions}
\end{figure}

The quality parameter $d$ was also used for the investigation of the ``fraction $N$'' discrimination parameter as a function of $N$. For this purpose for each run we calculated the quality parameter $d$ for the fast component integration window ($N$) in the range 10 to 100 ns. Results for the most important test chamber configurations and Xe concentrations are presented in figure~\ref{fig:psd_fractions}.
The variation of the window length from 40 to 100 ns does not affect significantly on the PSD quality parameter for the configurations with low Xe concentrations. This is in agreement with the previous result~\cite{lippincott2008scintillation}. At the same time, with increasing Xe concentration, the wide integration windows become less effective. This effect is caused by the decreasing of the slow component decay time and merging of the slow component and the fast component humps. The best separation between $\alpha$- and $\gamma$-particles in our case was reached with 40 ns integration window which is in agreement with ~\cite{peiffer2008pulse}. Thus, the F40 parameter can be used for all the test chamber configurations and Xe concentrations.
 
\begin{figure}[tbp]
\centering
\includegraphics[width=.45\textwidth]{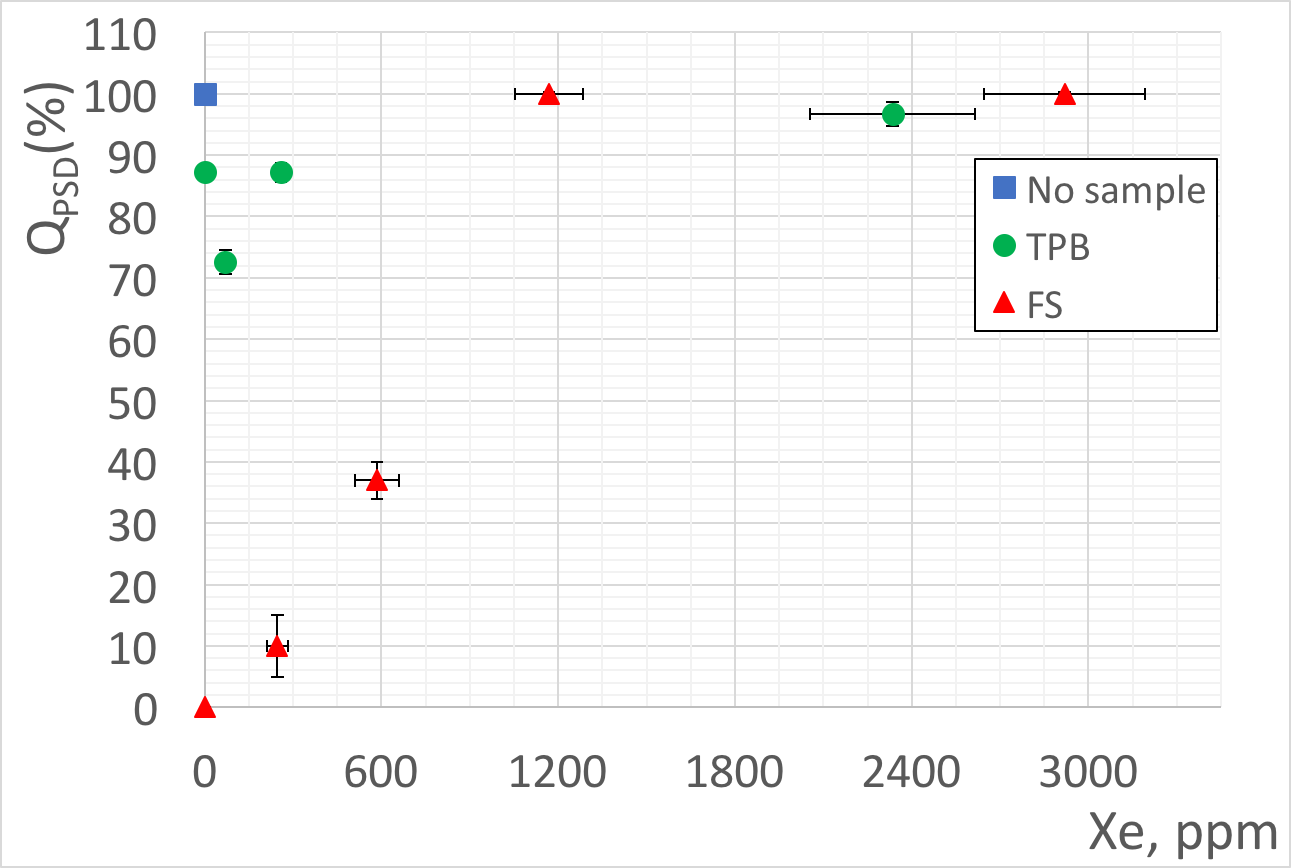}
\qquad
\includegraphics[width=.45\textwidth]{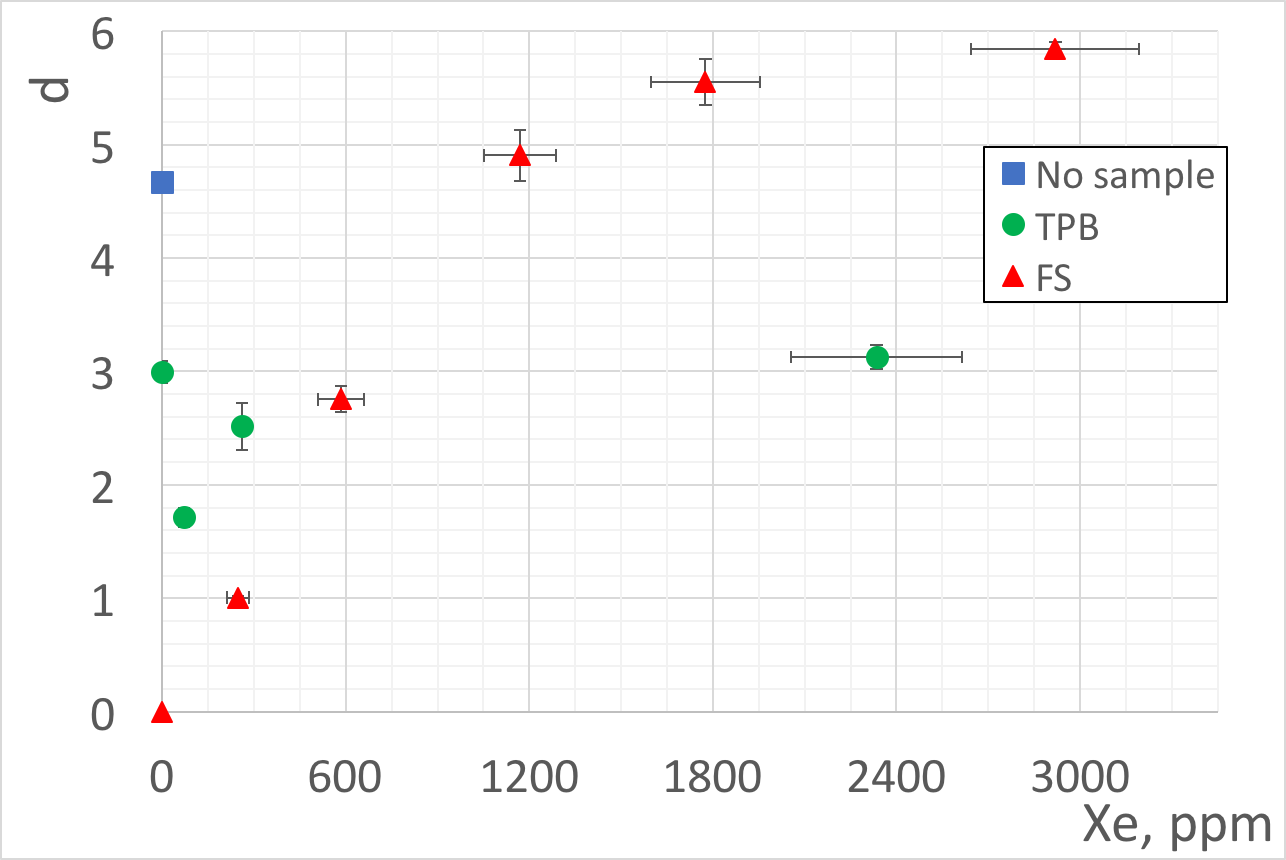}
\caption{\label{fig:Qpsd_figure} The dependence of quality parameters $Q_{PSD}$ (left) and $d$ (right) on Xe concentration; blue square --- test with no sample in front the PMT photocathode; green circles --- with TPB sample; red triangles --- with FS filter.}
\end{figure}

The dependence of both quality parameters ($Q_{PSD}$ and $d$) on Xe concentration is shown in figure~\ref{fig:Qpsd_figure}. The results obtained with the FS filter (red triangles in figure ~\ref{fig:Qpsd_figure}; see also figure~\ref{fig:main_result}) clearly 
demonstrate that the PSD capability of the LAr+Xe mixture is strongly related to the presence of the fast component in the scintillation signal. This relation was not evident in the previous studies with Xe-dopant because the fast component of LAr scintillation was re-emitted also by TPB.

Tests with the TPB sample (green circles in figure~\ref{fig:Qpsd_figure}) also show the decrease of the quality parameters with the decrease of Xe concentration. The similar behavior was reported in~\cite{wahl2014pulse}. However, we did not observe any PSD capability degradation at Xe concentration $\geq$ 300 ppm as proposed in~\cite{peiffer2008pulse}.  The decrease of the quality parameters at the low Xe concentrations is not so dramatic as in the previous case, because the fast component is still partly re-emitted by TPB. At zero Xe concentration, the $Q_{PSD}$ and $d$ are recovered since there is no energy transfer from the singlet $\mathrm{(^1\Sigma_u^+)}$ and triplet $\mathrm{(^3\Sigma_u^+)}$ excited states of $\mathrm{Ar_2^*}$ molecules to Xe.

One can see also that at the Xe concentration higher than $\sim$1000 ppm both quality parameters for the tests with FS filter reach the best values among all measurements. 

\subsection{Mixture stability}
\label{sec:MixtureStability}

Stability of the LAr+Xe mixture parameters in time is very important for long-term experiments. For low Xe concentrations was shown previously \cite{akimov2017study, himpslparticle} that the mixture remains stable in the long-term test. At the same time, there are statements \cite{raz1970experimental, neumeier2015intense} that Xe dopant may aggregate in LAr, and thus, reduce the effective Xe concentration inside the detector. The effect should be stronger at the higher Xe concentration. On the other hand, in \cite{yunker1960solubility} authors report about the good Xe solubility in LAr in concentrations up to several percent.

\begin{figure}[tbp]
\centering
\includegraphics[width=\textwidth]{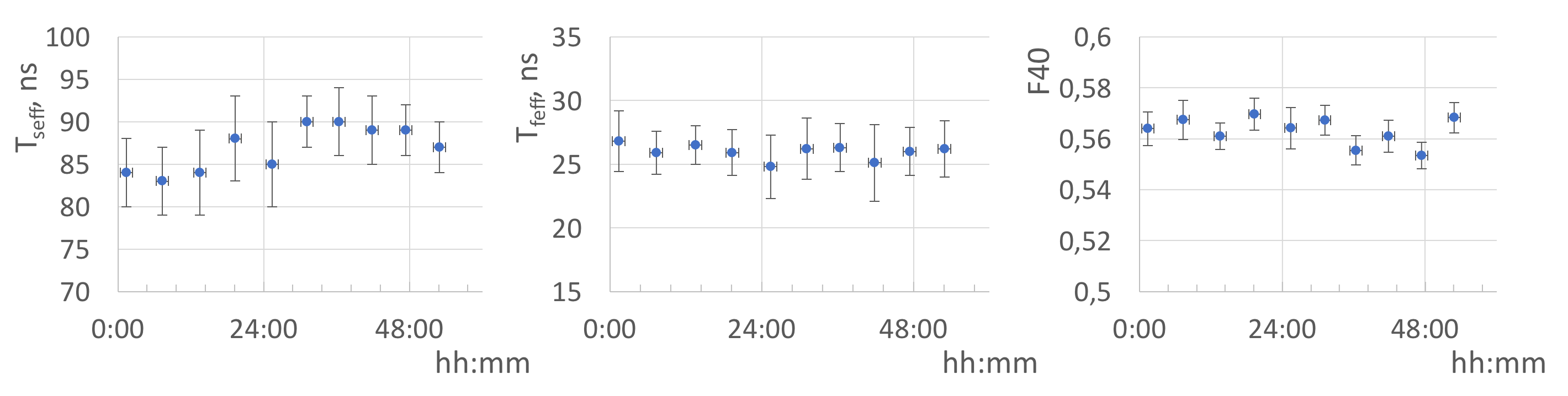}
\caption{\label{fig:mixture_stability} behavior in time of $T_{s_{eff}}$, $T_{f_{eff}}$, F40 parameters during 54-hour run.}
\end{figure}

In order to check the mixture long-term stability, a 54-hour continuous run was done at Xe concentration of $2920 \pm 270$ ppm. As shown above, there are several parameters (fast and slow decay time and the position of $\alpha$-band in PSD plot) which strongly depends on Xe concentration in LAr. The time behavior of $T_{s_{eff}}$, $T_{f_{eff}}$, F40 parameters are depicted in figure \ref{fig:mixture_stability}. The decay times of the fast $T_{f_{eff}}$ and slow $T_{s_{eff}}$ components represented in figure \ref{fig:mixture_stability} were obtained by a simplified procedure (denoted as $_{eff}$ due to this reason) instead of a multi-parameter fit with formulas \eqref{eq:Kubota_model}, \eqref{eq:Wahl_model}. A single-exponent fit was applied to the appropriate regions of the averaged waveform. This explains why the value of $T_{f_{eff}}$ shown in figure \ref{fig:mixture_stability} is significantly higher than that obtained by a multi-parameter fit. This approach is simpler than 4-term fit and has better sensitivity for the changes in Xe concentration, because the 4-term fit does not have any sensitivity to the changes in the fast component region with the increasing of Xe concentration as shown in section~\ref{sec:scintillation_parameters}.

In case of Xe aggregation or freezing on the detector walls, the $T_{s_{eff}}$ value should increase in time, in accordance with the power law behavior represented in figure \ref{fig:fast_and_slow_comp}. However, this parameter is sensitive to the change of Xe concentration only below 300 ppm (see figure \ref{fig:fast_and_slow_comp}). The $T_{f_{eff}}$ and F40 parameters are very sensitive for the variation of Xe concentrations in the region of $\sim$ 1000 ppm. Indeed, the fast part of the signal is faded out with the decrease of Xe concentration in this region. Eventually, the averaged waveform becomes one-hump shaped at the lower Xe concentration values. Thus, one should expect the increase of $T_{f_{eff}}$ value with the decrease of Xe concentration during the long run. The F40 parameter is strongly related to the presence of the fast component in the waveform, and obviously, it should decrease with the decrease of Xe concentration. These two parameters are sensitive to the changes in Xe concentration in the region below $\approx$1500 ppm since saturation of re-emission process is observed above this value. 

We have not observed any significant change of these parameters during 54-hour long continuous run. Unfortunately, we were unable to check the Xe concentration above 1500 ppm during the run. Thus, we can conclude that during the 54 hours of continuous operation at $\approx 94$ K the concentration of Xe in LAr did not drop below 1500 ppm.

There are also several things that are impossible to check in our small test chamber before introducing this WLS technique for bigger size LAr experiments. Namely, the uniformity of Xe distribution in vertical and radial directions and mixture stability are of interest for the large-size LAr detector where the temperature gradients are possible.

\section{Conclusion}
\label{sec:Conclusion}

We carried out a detailed study of the LAr scintillation re-emission with Xe-dopant as a single-stage WLS and when combined with TPB.

The main result of our investigation is that the fast component of LAr light is completely re-emitted by Xe at the high ($> 1500$ ppm) Xe concentrations in the LAr+Xe mixture. This was confirmed experimentally for the first time. 
We quantified the rate constant of the energy transfer for the fast component of LAr scintillation, which tends to be different from the rate constant for the slow component. This is in agreement with the overview \cite{buzulutskov2017photon} and the theoretical calculations of A.~Hitachi \cite{hitachi1993photon}, who proposed different rates of energy transfer to the Xe atoms for the fast and slow components.  However, the accuracy of our measurements is limited and this parameter should be studied in further experiments with higher time resolution and lower electronic noise.

The PSD capability is dependent on Xe concentration. It was demonstrated that it is strongly related to the fast component re-emission process. In the tests with high Xe concentrations with Xe-dopant as a single stage WLS, it was demonstrated that the PSD efficiency becomes better than either in the tests with the TPB sample combined with different Xe concentrations or in the tests with direct measurements of pure LAr.

The stability of mixture parameters for the highest tested Xe concentration of $\sim$3000 ppm was examined during a long (54 hours) continuous run for the first time. We observed no significant changes in any parameter related to Xe concentration: fast and slow effective decay time and the position of $\alpha$ band on the PSD plot. On the other hand, all these parameters are sensitive to the Xe concentration $< 1500$ ppm only and, thus, we can conclude that the effective Xe concentration did not fall below $\approx$1500 ppm. Here we are mostly in agreement with \cite{yunker1960solubility, wahl2014pulse, peiffer2008pulse, kubota1993suppression} who did not expect any problems with Xe solubility. We also did not observe any aggregation processes which can affect the obtained parameters, as proposed in \cite{neumeier2015intense, raz1970experimental}. Thus, we expect the Xe concentration to be stable for long-term experiments with Xe-doped LAr at temperatures around 94K.

\acknowledgments
The authors are thankful to the COHERENT collaboration for support of our studies, especially R.~Tayloe for inspiration of our investigations and for careful reading of this article. We also would like to say thanks to our colleagues from the RED collaboration for fruitful discussions.
We are grateful to S.Yu.~Semenov and A.G.~Fomin from the Laboratory of Physical and Chemical Research of FSE RTC RCSH  for the mass-spectrometric measurements of the Xe concentration in LAr+Xe mixture.
We would like to thank very much E. Bernard and A. Buzulutskov for interesting discussions of their papers  \cite{wahl2014pulse} and \cite{buzulutskov2017photon}, respectively.

\bibliographystyle{JHEP}
\bibliography{bib}

\end{document}